\documentclass{amsart}

\usepackage{amssymb,amsfonts}
\usepackage[all,arc]{xy}
\usepackage{enumerate}
\usepackage{mathrsfs}
\usepackage{amsrefs}
\usepackage{microtype}
\usepackage{multicol}
\usepackage{xcolor}
\usepackage{upgreek}
\usepackage[normalem]{ulem} 
\usepackage{tikz}
\usepackage{tikzscale}
\usepackage[figurename=Fig.]{caption}

\newtheorem{thm}{Theorem}[section]
\theoremstyle{definition}
\newtheorem{exmp}[thm]{Example}

\newcommand{\RR}{\mathbb{R}}
\newcommand{\bs}{\boldsymbol}

\makeatletter
\let\c@equation\c@thm
\makeatother
\numberwithin{equation}{section}

\begin{document}
\title[Computing MLEs for Gaussian Graphical Models in \texttt{Macaulay2}]{Computing Maximum Likelihood Estimates for Gaussian Graphical Models with Macaulay2}

\author{Carlos Am\'endola}
\address[Carlos Am\'endola]{Max Planck Institute for Mathematics in the Sciences, Leipzig, Germany}
\email{carlos.amendola@mis.mpg.de}

\author{Luis David Garc\'{\i}a Puente}
\address[Luis David Garc\'{\i}a Puente]{Department of Mathematics and Computer Science, Colorado College, Colorado Springs, CO, United States}
\email{lgarciapuente@coloradocollege.edu}

\author{Roser Homs}
\address[Roser Homs]{Department of Mathematics, Technical University of Munich, Germany}
\email{roser.homs@tum.de}

\author{Olga Kuznetsova}
\address[Olga Kuznetsova]{Department of Mathematics and Systems Analysis, Aalto University, P.O. Box 11100, FI-00076 Aalto, Finland }
\email{kuznetsova.olga@gmail.com}

\author{Harshit J Motwani}
\address[Harshit J Motwani]{Department of Mathematics: Algebra and Geometry, Ghent University, 9000 Gent, Belgium}
\email{harshitjitendra.motwani@ugent.be}


\begin{abstract}
We introduce the package \emph{GraphicalModelsMLE} for computing the maximum likelihood estimates (MLEs) of a Gaussian graphical model in the computer algebra system \emph{Macaulay2}. This package allows the computation of MLEs for the class of loopless mixed graphs. Additional functionality allows the user to explore the underlying algebraic structure of the model, such as its maximum likelihood degree and the ideal of score equations.
\end{abstract}

\maketitle

\markleft{}

\section{Introduction}

The purpose of the package \emph{GraphicalModelsMLE}\footnote{\emph{GraphicalModelsMLE} version 0.3 is included in \emph{Macaulay2} version 1.17, the revised \emph{GraphicalModelsMLE} version 1.0 is available at \url{https://github.com/roserhp/GraphicalModelsMLE} and is expected to appear in \emph{Macaulay2} version 1.20.} is to extend the functionality of \emph{Macaulay2} \cite{grayson2002macaulay2} related to algebraic statistics, specifically allowing computations of maximum likelihood estimates of Gaussian graphical models. While \emph{GraphicalModels} is an existing package that already provides useful information such as conditional independence ideals and vanishing ideals for such models, the fundamental statistical inference task of computing maximum likelihood estimates is missing. This package aims to fill this void and also extend the functionality of \emph{GraphicalModels} to handle more general types of graphs, in particular, \emph{loopless mixed graphs} (LMG). This class of graphs was introduced by Sadeghi and Lauritzen in~\cite{sadeghi2014markov} in order to unify the Markov theory of several classical types of graphs such as undirected graphs, directed acyclic graphs, summary graphs and ancestral graphs~\cite{lauritzen1996}.

The algebraic framework of \emph{Macaulay2} permits us to use both commutative algebra and numerical algebraic geometry to obtain a guaranteed global optimal solution by computing all critical points of the log-likelihood function. This is  different from the classical statistical approach of the \emph{R} package \emph{ggm} \cite{ggmR}, and more in line with the recent numerical algebraic geometry approach from the package \emph{LinearCovarianceModels.jl} in \emph{Julia} \cite{LinearCovarianceModels}. The package \emph{GraphicalModelsMLE} is a complement to these two, handling some Gaussian graphical models not covered by them (\emph{LinearcovarianceModels.jl} version 0.2 and \emph{ggm} version 2.5). The capabilities of our package are limited by the feasability of the Gr\"obner basis computations involved.

Given a data sample of $n$ independent and identically distributed random vectors  $X^{(1)}$, $\dots$, $X^{(n)}$ that follow an $m$-dimensional multivariate Gaussian distribution $\mathcal{N}(\mu,\Sigma)$, the \emph{maximum likelihood estimate} (MLE) for the covariance matrix $\Sigma$ is the matrix that best explains the observed data, in the sense that it maximizes the likelihood function of the Gaussian model (see Section~\ref{sec3}).

\section{Graphical models of loopless mixed graphs}\label{sec2}

A \emph{mixed graph} $G=(V,E)$ is a graph with undirected edges $i-j$, directed edges $i\to j$ and bidirected edges $i \leftrightarrow j$. A \emph{directed cycle} is a cycle formed by directed edges after identifying the vertices that are connected by undirected or bidirected edges. A \emph{loopless mixed graph} (LMG) is a mixed graph without loops or directed cycles. We allow double edges of the types directed-undirected and directed-bidirected. See Figures~\ref{fig:4-cycle} and \ref{fig:mixed graph} for examples and Figure~\ref{fig:multi-edge-cycle} for a non-example. 

\begin{figure}[h]

    \centering
\tikzset{every picture/.style={line width=0.75pt}} 

\begin{minipage}[t]{.33\textwidth}
\captionsetup{width=0.8\textwidth, format=hang}
\centering

\includegraphics{4-cycle}
    \captionof{figure}{Undirected graph} \label{fig:4-cycle}
\end{minipage}
\begin{minipage}[t]{.33\textwidth}
\captionsetup{width=0.8\textwidth, format=hang}
\centering
\includegraphics{mixed-graph}
    \captionof{figure}{Mixed graph with no directed cycles} \label{fig:mixed graph}
\end{minipage}
\begin{minipage}[t]{.33\textwidth}
\captionsetup{width=0.8\textwidth, format=hang}
\centering
\includegraphics{directed-cycle}
    \captionof{figure}{Mixed graph with directed cycles} \label{fig:multi-edge-cycle}
\end{minipage}
\end{figure}

Following~\cite{sullivant2010trek}, we assume that the nodes of $G$ are partitioned as $V = U\cup W$, such that:
\begin{itemize}
\item if $i-j$ in $G$ then $i,j\in U$
\item if $i\leftrightarrow j$ in $G$ then $i,j\in W$ 
\item there is no directed edge $i\to j$ in $G$ such that $i\in W$ and $j\in U$.
\end{itemize}

Our definition differs from the one in~\cite{sadeghi2014markov} in that we do not allow multiple edges of the same type, which is due to the setup of the \emph{Graphs} package. Also note that the partition of vertices excludes multiple edges of the type undirected-bidirected. In addition, we prohibit directed cycles, which ensures there is an ordering on the vertices such that all vertices in $U$ come before vertices in $W$, and whenever $i\to j$ we have $i<j$. For simplicity of the algorithm, in our \emph{Macaulay2} implementation we will require the graph to be provided with such an ordering of the vertices (see more details in the description of \texttt{partitionLMG} in Section \ref{sec6}).

A Gaussian graphical model imposes constraints on the covariance matrix of a Gaussian distribution. More precisely, a loopless mixed graph $G=(V,E)$ gives rise to the space of covariance matrices $\Sigma \in \RR^{|V| \times |V|}$ of the form~\cite{sullivant2010trek}*{Section 2.3}
 \begin{equation}\label{eq:parametrization}
\Sigma\quad=\quad (I-\Lambda)^{-T}\left[\begin{array}{cc}
K^{-1} & \bs 0\\
\bs 0 & \Psi
\end{array}
\right](I-\Lambda)^{-1},
\end{equation}
where 
\begin{itemize}
\item[(i)] $\Lambda=[\lambda_{ij}]\in \RR^{|V|\times |V|}$ is such that $\lambda_{ij}=0$ whenever $i\to j \notin E$;
\item[(ii)] $K=[k_{ij}] \in \RR^{|U|\times |U|}$ is symmetric positive definite such that $k_{ij}=0$ whenever $i-j\notin E$;
\item[(iii)] $\Psi =[\psi_{ij}]\in \RR^{|W|\times |W|}$ is symmetric positive definite such that  $\psi_{ij}=0$ whenever $i\leftrightarrow j\notin E$.
\end{itemize}

\section{Maximum likelihood estimates}\label{sec3}
Let the data sample consist of $n$ independent identically distributed random vectors $X^{(1)}, \dots, X^{(n)}$ sampled from an $m$-dimensional Gaussian distribution $ \mathcal{N}_m(\mu,\Sigma)$. The parameter space of the corresponding statistical model is $\Theta=\mathbb{R}^m \times\Theta_2\subseteq\mathbb{R}^m \times PD_m$, where $\Theta_2$ is the space of covariance matrices $\Sigma$ and $PD_m$ is the cone of $m \times m$ positive-definite matrices. The maximum likelihood estimates for the covariance matrix is determined by maximizing the log-likelihood function
\begin{equation}\label{eq: solverMLE obj}
\ell(\Sigma)=-\frac{n}{2}\log \det \Sigma - \frac{n}{2}\operatorname{tr}\left(S\Sigma^{-1}\right)
\end{equation}
over $\Sigma \in \Theta_2$~\cite{sullivant2018algebraic}*{Proposition 7.1.9}, where $S$ is the sample covariance matrix. 
The function \texttt{solverMLE}  allows to compute this optimum when $\Theta_2$ is induced by \eqref{eq:parametrization}. It does so by calculating the critical points of the log-likelihood function and selecting the points corresponding to the maximum value in the cone of positive definite matrices. The default output is the maximum value of $\ell(\Sigma)$, the list of maximum likelihood estimates for the covariance matrix and the maximum likelihood degree of the model. 

For undirected graphs, the MLE for the covariance matrix is known to be the unique positive definite critical point of the likelihood function. In particular, it is a 
positive definite matrix completion to the partial sample covariance matrix. 
See~\cite{Uhler}*{Theorem 2.1} or~\cite{OWL}*{Theorem 2.1.14} for more details.

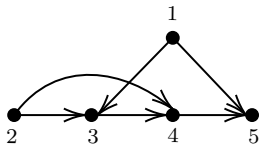
\begin{figure}[h!]
\centering
\tikzset{every picture/.style={line width=0.75pt}} 

\begin{tikzpicture}[x=0.75pt,y=0.75pt,yscale=-1,xscale=1]

\draw  [fill={rgb, 255:red, 0; green, 0; blue, 0 }  ,fill opacity=1 ] (113.5,31.5) .. controls (113.5,29.84) and (112.16,28.5) .. (110.5,28.5) .. controls (108.84,28.5) and (107.5,29.84) .. (107.5,31.5) .. controls (107.5,33.16) and (108.84,34.5) .. (110.5,34.5) .. controls (112.16,34.5) and (113.5,33.16) .. (113.5,31.5) -- cycle ;

\draw  [fill={rgb, 255:red, 0; green, 0; blue, 0 }  ,fill opacity=1 ] (33.5,70.5) .. controls (33.5,68.84) and (32.16,67.5) .. (30.5,67.5) .. controls (28.84,67.5) and (27.5,68.84) .. (27.5,70.5) .. controls (27.5,72.16) and (28.84,73.5) .. (30.5,73.5) .. controls (32.16,73.5) and (33.5,72.16) .. (33.5,70.5) -- cycle ;

\draw  [fill={rgb, 255:red, 0; green, 0; blue, 0 }  ,fill opacity=1 ] (113.5,70.5) .. controls (113.5,68.84) and (112.16,67.5) .. (110.5,67.5) .. controls (108.84,67.5) and (107.5,68.84) .. (107.5,70.5) .. controls (107.5,72.16) and (108.84,73.5) .. (110.5,73.5) .. controls (112.16,73.5) and (113.5,72.16) .. (113.5,70.5) -- cycle ;

\draw  [fill={rgb, 255:red, 0; green, 0; blue, 0 }  ,fill opacity=1 ] (72.5,70.5) .. controls (72.5,68.84) and (71.16,67.5) .. (69.5,67.5) .. controls (67.84,67.5) and (66.5,68.84) .. (66.5,70.5) .. controls (66.5,72.16) and (67.84,73.5) .. (69.5,73.5) .. controls (71.16,73.5) and (72.5,72.16) .. (72.5,70.5) -- cycle ;

\draw  [fill={rgb, 255:red, 0; green, 0; blue, 0 }  ,fill opacity=1 ] (153.5,70.5) .. controls (153.5,68.84) and (152.16,67.5) .. (150.5,67.5) .. controls (148.84,67.5) and (147.5,68.84) .. (147.5,70.5) .. controls (147.5,72.16) and (148.84,73.5) .. (150.5,73.5) .. controls (152.16,73.5) and (153.5,72.16) .. (153.5,70.5) -- cycle ;

\draw    (30.5,70.5) -- (64.5,70.5) ;
\draw [shift={(66.5,70.5)}, rotate = 180] [color={rgb, 255:red, 0; green, 0; blue, 0 }  ][line width=0.75]    (10.93,-3.29) .. controls (6.95,-1.4) and (3.31,-0.3) .. (0,0) .. controls (3.31,0.3) and (6.95,1.4) .. (10.93,3.29)   ;

\draw    (110.5,31.5) -- (73.9,69.07) ;
\draw [shift={(72.5,70.5)}, rotate = 314.26] [color={rgb, 255:red, 0; green, 0; blue, 0 }  ][line width=0.75]    (10.93,-3.29) .. controls (6.95,-1.4) and (3.31,-0.3) .. (0,0) .. controls (3.31,0.3) and (6.95,1.4) .. (10.93,3.29)   ;

\draw    (110.5,31.5) -- (146.12,69.05) ;
\draw [shift={(147.5,70.5)}, rotate = 226.51] [color={rgb, 255:red, 0; green, 0; blue, 0 }  ][line width=0.75]    (10.93,-3.29) .. controls (6.95,-1.4) and (3.31,-0.3) .. (0,0) .. controls (3.31,0.3) and (6.95,1.4) .. (10.93,3.29)   ;

\draw    (69.5,70.5) -- (105.5,70.5) ;
\draw [shift={(107.5,70.5)}, rotate = 180] [color={rgb, 255:red, 0; green, 0; blue, 0 }  ][line width=0.75]    (10.93,-3.29) .. controls (6.95,-1.4) and (3.31,-0.3) .. (0,0) .. controls (3.31,0.3) and (6.95,1.4) .. (10.93,3.29)   ;

\draw    (113.5,70.5) -- (145.5,70.5) ;
\draw [shift={(147.5,70.5)}, rotate = 180] [color={rgb, 255:red, 0; green, 0; blue, 0 }  ][line width=0.75]    (10.93,-3.29) .. controls (6.95,-1.4) and (3.31,-0.3) .. (0,0) .. controls (3.31,0.3) and (6.95,1.4) .. (10.93,3.29)   ;

\draw    (30.5,70.5) .. controls (52.06,40.61) and (87.07,47.7) .. (109.16,66.34) ;
\draw [shift={(110.5,67.5)}, rotate = 221.55] [color={rgb, 255:red, 0; green, 0; blue, 0 }  ][line width=0.75]    (10.93,-3.29) .. controls (6.95,-1.4) and (3.31,-0.3) .. (0,0) .. controls (3.31,0.3) and (6.95,1.4) .. (10.93,3.29)   ;

\draw (106.1,13.92) node [anchor=north west][inner sep=0.75pt]  [font=\footnotesize,rotate=-0.79] [align=left] {1};

\draw (25.1,76.03) node [anchor=north west][inner sep=0.75pt]  [font=\footnotesize,rotate=-0.79] [align=left] {2};

\draw (66.1,76.03) node [anchor=north west][inner sep=0.75pt]  [font=\footnotesize,rotate=-0.79] [align=left] {3};

\draw (106.46,75.53) node [anchor=north west][inner sep=0.75pt]  [font=\footnotesize,rotate=-0.79] [align=left] {4};

\draw (147.1,76.03) node [anchor=north west][inner sep=0.75pt]  [font=\footnotesize,rotate=-0.79] [align=left] {5};

\end{tikzpicture}

\caption{Verma graph} \label{fig:DAG}

\end{figure}

\begin{exmp}\label{DAG}
We consider the directed acyclic graph $G$ known as the Verma graph, see Figure~\ref{fig:DAG} and \cite{OWL}*{Example 3.3.14}. We take as sample data the columns of a real matrix $U$ generated with the command \texttt{random} in \emph{Macaulay2} and compute the MLE for the covariance matrix that best explains the data within the graphical model given by $G$.
\footnotesize{
\begin{verbatim}
i1: loadPackage "GraphicalModelsMLE";
i2: G=digraph{{1,3},{1,5},{2,3},{2,4},{3,4},{4,5}};
i3: U=matrix {{.0137595, .983763, .963969, .152094, .0453326},
              {.527344, .597575, .777622, .97937, .112339}, 
              {.097922, .300712, .333058, .824002, .420228},
              {.849322, .594136, .114729,.69734, .98773},
              {.764547, .42209, .480193, .246573, .846734}};
i4: solverMLE(G,U)
o4 = (8.77485, | .115729     -6.32685e-18 -.0387187 .00115181 .102733   |, 1)
               | 6.10884e-18 .053294      .0392544  -.0356783 .00701454 |
               | -.0387187   .0392544     .0807822  -.0278223 -.0289767 |
               | .00115181   -.0356783    -.0278223 .105095   -.0196375 |
               | .102733     .00701454    -.0289767 -.0196375 .148723   |
\end{verbatim}}
\end{exmp}

\begin{exmp}\label{4-cycle} We compute the MLE for the covariance matrix of the graphical model associated to the undirected 4-cycle, see Figure~\ref{fig:4-cycle}. We encode the sample data by a matrix $U$ generated as in Example \ref{DAG} and compute the sample covariance matrix $S=\frac{1}{n}UU^T$.
 
\footnotesize{
\begin{verbatim}
i1 : loadPackage "GraphicalModelsMLE";    
i2 : G=graph{{1,2},{2,3},{3,4},{4,1}};
i3 : S=matrix {{.105409, -.0745495, -.0186132, .0621907},
               {-.0745495, .0783734,-.00844503,-.0872842},
               {-.0186132, -.00844503, .128307, .0230245}, 
               {.0621907, -.0872842, .0230245,.109849}};
i4 : solverMLE(G,S,SampleData=>false)
o4 = (6.62005, | .105409   -.0745495  .0124099   .0621907  |, 5)
               | -.0745495 .0783734   -.00844503 -.0439427 |
               | .0124099  -.00844503 .128307    .0230245  |
               | .0621907  -.0439427  .0230245   .109849   |
\end{verbatim}}
\normalsize{
Note that all entries in the MLE for the covariance matrix coincide with the entries in the sample covariance matrix except for those corresponding to non-edges of the graph. See \cite{NonLinearAlgebra}*{Example 12.16} for more on a positive definite matrix completion problem associated to the 4-cycle.}

\end{exmp}

For more general types of graphs, uniqueness of the positive definite critical points is no longer guaranteed. In the mixed graph below, the optimization problem has a global maximum, but there are also local maxima, see Example \ref{2LocalMaxima}. 
    
\begin{exmp}\label{mixed graph} We compute the MLE for the covariance matrix of the graphical model associated to the loopless mixed graph with undirected edge $1-2$, directed edges $1\rightarrow 3,2\rightarrow 4$ and bidirected edge $3\leftrightarrow 4$, see Figure~\ref{fig:mixed graph}. $S$ is a sample covariance matrix computed from sample data encoded in a rational matrix obtained again with the command \texttt{random}.
\footnotesize{
\begin{verbatim}
i2 : G = mixedGraph(graph{{1,2}},digraph{{1,3},{2,4}},bigraph{{3,4}});
i3 : S=matrix {{34183/50000, 716539/10000000, 204869/250000, 12213/25000}, 
               {716539/10000000, 112191/500000, 309413/1000000, 1803/4000}, 
               {204869/250000, 309413/1000000, 3849/3125, 15172/15625}, 
               {12213/25000, 1803/4000, 15172/15625, 4487/4000}};
i4 : solverMLE(G,S,SampleData=>false)
o4 = (9.36624, {| .68366   .0716539 1.00282   .234375   |}, 5)
                | .0716539 .224382  .105105   .733937   |
                | 1.00282  .105105  1.76955   -.0700599 |
                | .234375  .733937  -.0700599 2.97432   |
\end{verbatim}}
\end{exmp}

\section{Ideal of score equations}\label{sec4}
The critical points of the log-likelihood function $\ell(\Sigma)$ are the solutions to the system of equations obtained by taking partial derivatives of $\ell$ with respect to all variables in the entries of $\Sigma$ from our construction in~(\ref{eq:parametrization}) and setting them to zero:
\begin{equation}\label{eq:scoreEquations}
-\frac{\partial}{\partial (\cdot)} \det \Sigma-\det \Sigma\frac{\partial }{\partial (\cdot)}\operatorname{tr(S \Sigma^{-1})}=0,
\end{equation}
where $\frac{\partial}{\partial (\cdot)}$ stands for the partial derivatives with respect to the variables $\lambda_{ij},k_{ij},\psi_{ij}$ in the entries of the covariance matrix $\Sigma$ from (\ref{eq:parametrization}). These polynomial equations are called \emph{score equations}. The command \texttt{scoreEquations} returns the ideal generated by such polynomials, which lives in the polynomial ring $\mathbb{Q}[\lambda_{ij},k_{ij},\psi_{ij}]$. From an algebraic perspective, this ideal is already of interest on its own, see \cite{sullivant2018algebraic}*{Chapter 7}.

Note that the log-likelihood function depends both on the sample covariance matrix and the graphical model. Therefore our implementation of \texttt{scoreEquations} requires as input the sample data along with information about the model. The latter is obtained via the command \texttt{gaussianRing} in the package \emph{GraphicalModels} (see \cite{garcia2013graphical} and examples in Section \ref{sec6}), which produces a ring associated to the graph $G$ that stores all relevant features of the graphical model.

\begin{exmp}\label{0dim}
We compute the ideal of score equations associated to the 4-cycle after creating the graph $G$ as in Example \ref{4-cycle}. We now consider as input data the sample data encoded in the columns of the integer matrix $U$ below, obtained via the command \texttt{random}.
\begin{verbatim}
i5 : U=matrix{{3,5,9,5},{1,6,1,5},{2,9,6,6},{2,5,0,4}};
i6 : J=scoreEquations(gaussianRing G,U);
o6 : Ideal of QQ[k   , k   , k   , k   , k   , k   , k   , k   ]
                  1,1   2,2   3,3   4,4   1,2   1,4   2,3   3,4
i7 : dim J
o7 = 0
\end{verbatim}
The ideal of score equations $J$ is generated by 14 non-homogeneous polynomials in $\mathbb{Q}[k_{1,1},k_{1,2},k_{1,4}, k_{2,2}, k_{2,3}, k_{3,3}, k_{3,4}, k_{4,4}]$: 4 linear polynomials and 10 quadratic polynomials such as $1312002k_{3,4}^2-387081k_{1,2}+109860k_{1,4}+1972025k_{2,3}-898518k_{3,4}-291556$. Since this ideal is zero-dimensional, the log-likelihood function  $\ell(\Sigma)$ defined in~(\ref{eq: solverMLE obj}) has finitely many complex critical points, as will be discussed in Section~\ref{sec5}.
\end{exmp}    

\begin{exmp}\label{2LocalMaxima} We want to obtain all local maxima of the log-likelihood function associated to the graphical model in Example \ref{mixed graph}. We write $\lambda$ as $l$ and $\psi$ as $p$ in the code. 
The score equations generate an ideal in $\mathbb{Q}[k_{1,1},k_{2,2},k_{1,2},l_{1,3},l_{2,4},p_{3,3},p_{4,4},p_{3,4}]$ and we display their solutions in the \emph{Macaulay2} session below.
We retrieve the covariance matrix $\Sigma$ with rational entries in variables $k_{1,1},k_{2,2},k_{1,2},l_{1,3},l_{2,4},p_{3,3},p_{4,4},p_{3,4}$ using the optional output \texttt{CovarianceMatrix} in \texttt{scoreEquations}.

\footnotesize{
\begin{verbatim}
i5 : R = gaussianRing G;
i6 : (J,Sigma)=scoreEquations(R,S,SampleData=>false,CovarianceMatrix=>true);
i7 : dim J, degree J
o7 = (0, 5)
i8 : sols=zeroDimSolve(J);netList sols
     +--------------------------------------------------------------------------+
o8 = |{1.51337, 4.61101, -.483277, 1.46684, 3.27093, .298576, .573665, -.41385} |
     +--------------------------------------------------------------------------+
     |{1.51337, 4.61101, -.483277, 1.39884+.440525*ii, 2.45466-.923165*ii,      |
     |.144129+.120574*ii,.0696297-.184692*ii,-.19668+.0553853*ii}               |
     +--------------------------------------------------------------------------+
     |{1.51337, 4.61101, -.483277, 1.39884-.440525*ii, 2.45466+.923165*ii,      |
     |.144129-.120574*ii, .0696297+.184692*ii, -.19668-.0553853*ii}             |
     +--------------------------------------------------------------------------+
     |{1.51337, 4.61101, -.483277, .684147, .979681, .430388, .453924, .381688} |
     +--------------------------------------------------------------------------+
     |{1.51337, 4.61101, -.483277, .988484, 1.64649, .279607, .245722, .0952865}|
     +--------------------------------------------------------------------------+

\end{verbatim}}

\noindent 
\normalsize{How many of the 3 real critical points correspond to positive definite matrices that are local maxima of the log-likelihood function? We first check that they correspond to positive definite matrices by substituting the three real solutions in the covariance matrix $\Sigma$.}

\footnotesize{
\begin{verbatim}
i9 : checkPD(apply(sols,i->sub(Sigma,matrix{coordinates(i)})))
o9 = |.68366   .0716539 1.00282   .234375   |,| .68366   .0716539 .467724  .070198 |, 
     |.0716539 .224382  .105105   .733937   | | .0716539 .224382  .0490218 .219823 | 
     |1.00282  .105105  1.76955   -.0700599 | | .467724  .0490218 .75038   .429714 | 
     |.234375  .733937  -.0700599 2.97432   | | .070198  .219823  .429714  .66928  |  
                       | .68366   .0716539 .675787  .117978 |
                       | .0716539 .224382  .0708287 .369443 |
                       | .675787  .0708287 .947611  .211905 |
                       | .117978  .369443  .211905  .854009 |
\end{verbatim}}

\noindent
\normalsize{The MLE for the covariance matrix obtained in Example \ref{mixed graph} corresponds to the first positive definite matrix in the list above. The eigenvalues of the Hessian matrix computed below tell us which kind of critical point we have for each of the 3 real solutions --- for a discussion about the properties of positive-semidefinite matrices see~\cite{blekherman2012semidefinite}*{Appendix A}.}

\footnotesize{
\begin{verbatim}
-- compute Jacobian matrix (i.e. score equations)
i10 : scoreEq=-1/det Sigma*jacobianMatrixOfRationalFunction(det Sigma)-
jacobianMatrixOfRationalFunction(trace(S*(inverse Sigma)));
-- compute Hessian matrix
i11 : Hessian=matrix for f in flatten entries scoreEq list 
flatten entries jacobianMatrixOfRationalFunction(f);
-- compute eigenvalues of the Hessian matrix evaluated at real points in sols  
i12 : apply({sols_0,sols_3,sols_4},i->eigenvalues sub(Hessian,matrix{coordinates(i)}))
o12 = {{-.516478   }, {-.516478 }, {-.516478 }}
       {-.271913   }  {-.271913 }  {-.271913 }
       {-.0464172  }  {-.0464172}  {-.0464172}
       {-9869730000}  {-414.15  }  {-59.7135 }
       {-128887    }  {-28.6352 }  {1.52598  }
       {-58261.1   }  {-6.1936  }  {-2.80504 }
       {-773.513   }  {-1.64689 }  {-11.5533 }
\end{verbatim}}

\noindent
\normalsize{The first two points are local maxima and the last point is a saddle point. This shows that the log-likelihood function of this model is not a concave function, see \cite{Drton}}.
\end{exmp}

\begin{figure}
    \centering
\tikzset{every picture/.style={line width=0.75pt}} 
\includegraphics{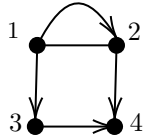}
    \captionof{figure}{LMG with two types of edges between $1$ and $2$} \label{fig:multi-edge}
\end{figure}

\begin{exmp}\label{multi-edge}
 Next we compute the ideal of score equations associated to a mixed graph that has two different types of edges connecting the same two vertices: directed edges $1\rightarrow 3,1\rightarrow 2,2\rightarrow 4,3\rightarrow 4$ and undirected edge $1-2$, see Figure~\ref{fig:multi-edge}. 
\begin{verbatim}
i2 : G = mixedGraph(digraph{{1,3},{1,2},{2,4},{3,4}},graph{{1,2}});
i3 : R = gaussianRing G;
i4 : U = random(RR^4,RR^4);
i5 : J=scoreEquations(R,U);
i6 : dim J
o6 = 1
\end{verbatim}
 Note that in this case, as opposed to Example \ref{0dim}, the log-likelihood function (\ref{eq: solverMLE obj}) has infinitely many complex critical points. Since our package evaluates the objective function in all critical points, in such a scenario the MLE cannot be computed.
\end{exmp}

\section{Maximum likelihood degree}\label{sec5}

The \emph{maximum likelihood degree} (ML degree) of a model is defined as the number of complex critical points of the log-likelihood function $\ell(\Sigma)$ from~(\ref{eq: solverMLE obj}) for generic sample data, see \cite{sullivant2018algebraic}*{Definition 7.1.4}. For a more algebraic flavour of the notion of ML degree, see \cite{MichalekSturmfels}*{Definition 5.4}.

Note that the ML degree is only well-defined when the ideal of score equations is zero-dimensional. A typical way where this fails is where the model becomes non-identifiable. See, for example, \cite{structurelearning} for some sufficient conditions to avoid non-identifiability and preservation of dimension of the model in terms of the number of parameters.

It is important to observe that for generic data the solutions to score equations are all distinct, see \cite{amendola2020}*{Remark 2.1,\,Lemma 2.2}. Computing the algebraic degree of the zero-dimensional score equations ideal via the \texttt{degree} function in \emph{Macaulay2} is equivalent to computing the number of complex solutions - without multiplicity - to the score equations~(\ref{eq:scoreEquations}).

In our implementation of the \texttt{MLdegree} function in \emph{Macaulay2} a random sample matrix is used as sample data. Therefore, the ML degree of the graphical model we provide is correct with probability 1.

\begin{exmp} The ML degree of the 4-cycle 
can be directly computed as follows:
\begin{verbatim}
i2 : G=graph{{1,2},{2,3},{3,4},{4,1}};
i3 : MLdegree(gaussianRing G)
o3 = 5
\end{verbatim}
\end{exmp}    

In the case of ideals of score equations with positive dimension, \texttt{MLdegree} will still compute the degree of the ideal but this no longer matches the number of solutions to the score equations.

\begin{exmp} Continuing with Example \ref{multi-edge}, where the ideal of score equations is 1-dimensional, \texttt{MLdegree} does not provide a meaningful answer.
\begin{verbatim}
i2: G=mixedGraph(digraph{{1,3},{1,2},{2,4},{3,4}},graph{{1,2}});
i3: MLdegree(gaussianRing G)     
error: the ideal of score equations has dimension 1 > 0, 
so ML degree is not well-defined. The degree of this ideal is 2.
\end{verbatim}
\end{exmp}
 
\section{Updates in related packages}\label{sec6}

\emph{GraphicalModelsMLE} relies on the new package \emph{StatGraphs} 0.1 and the updated packages \emph{Graphs} 0.3.3 and \emph{GraphicalModels} 2.0 (see \cite{garcia2013graphical} for version 1.0).

We created a dedicated package \emph{StatGraphs} for graph theoretic functions relevant to algebraic statistics. It contains the functions \texttt{isCyclic}, \texttt{isSimple}, \texttt{isLoopless}  and \texttt{partitionLMG} to deal with loopless mixed graphs.

The function \texttt{partitionLMG} computes the partition $V=U\cup W$ of vertices of a loopless mixed graph described in Section \ref{sec2}. Vertices in the input graph need to be ordered such that (1) all vertices in $U$ come before vertices in $W$ and (2) if there is a directed edge $i\rightarrow j$, then $i<j$.

\begin{exmp}\label{ex:partitionLMG} The vertices of the loopless mixed graph in Example \ref{mixed graph} 
are partitioned into $U=\lbrace 1,2\rbrace$ and $W=\lbrace 3,4\rbrace$.

\footnotesize{\begin{verbatim}
i1 : loadPackage "StatGraphs";	   
i2 : G = mixedGraph(digraph {{1,3},{2,4}},bigraph{{3,4}},graph{{1,2}});
i3 : partitionLMG G
o3 = ({1, 2}, {3, 4})
o3 : Sequence
\end{verbatim}}
\end{exmp}
The central object in the implementation of our MLE algorithm is \texttt{gaussianRing} from the package \emph{GraphicalModels}. 

\begin{exmp}\label{ex:gaussianRing} We compute the ring associated to the graph in Example \ref{ex:partitionLMG} and display the variables of
the ring as entries of matrices. We write $\lambda$ as $l$ and $\psi$ as $p$ in the code. 
\begin{multicols}{2}
\footnotesize{
\begin{verbatim}
i4 : loadPackage "GraphicalModels";
i5 : R=gaussianRing G;
i6 : undirectedEdgesMatrix R	   
o6 = | k_(1,1) k_(1,2) |
     | k_(1,2) k_(2,2) |
i7 : directedEdgesMatrix R
o7 = | 0 0 l_(1,3) 0       |
     | 0 0 0       l_(2,4) |
     | 0 0 0       0       |
     | 0 0 0       0       |
\end{verbatim}
\columnbreak
\begin{verbatim}
i8 : bidirectedEdgesMatrix R
o8 = | p_(3,3) p_(3,4) |
     | p_(3,4) p_(4,4) |
i9 : covarianceMatrix R
o9 = | s_(1,1) s_(1,2) s_(1,3) s_(1,4) |
     | s_(1,2) s_(2,2) s_(2,3) s_(2,4) |
     | s_(1,3) s_(2,3) s_(3,3) s_(3,4) |
     | s_(1,4) s_(2,4) s_(3,4) s_(4,4) |
\end{verbatim}}
\end{multicols}
\end{exmp}

In version 2.0 of \emph{GraphicalModels}, we updated the functionalities of the method \texttt{gaussianRing} -- and its related methods -- in order to accept loopless mixed graphs with undirected, directed and bidirected edges.
           
Note that mixed graphs that include undirected edges are required to have an ordering compatible with \texttt{partitionLMG}. For mixed graphs with only directed and bidirected edges this is no longer necessary, as in version 1.0 of \emph{GraphicalModels}.

\subsection*{Acknowledgements}

We thank Michael Stillman and Daniel Grayson for their support
and help with the inner workings of \emph{Macaulay2} during the 2020 online Warwick \emph{Macaulay2} Workshop. We are also grateful to Elina Robeva and David Swinarski for creating the first draft of \emph{GraphicalModelsMLE}, and to Piotr Zwiernik for useful discussions.
Harshit J Motwani was partially supported by UGent BOF/ STA/201909/038 and FWO grant G0F5921N.
\bibliographystyle{alpha}
\bibliography{bibliography}

\end{document}